\documentclass{epl}
\usepackage{epsfig}
\usepackage{amssymb,amsbsy,amsmath}
\newtheorem{theorem}{Theorem}

\newcommand{\op}[1]{%
    \fontdimen12\textfont3=2pt\fontdimen12\scriptfont3=1.4pt%
    \!\null\mathop{\vphantom{#1}\smash{#1}}\limits_{\sim}\null\!}

\def\bra#1{\langle \, {#1} \, | \,}
\def\ket#1{\, | \, {#1} \, \rangle}
\newcommand{\Tr}[1]{\text{Tr}\left\{#1\right\}}
\title{Approximating parabolas as natural bounds of Heisenberg spectra:
Reply on the comment of O. Waldmann}
\shorttitle{Reply}
\author{Heinz-J\"urgen Schmidt\inst{1} \and J\"urgen Schnack\inst{1}, \and Marshall Luban\inst{2}}
\shortauthor{H.-J.~Schmidt \and J.~Schnack \and M.~Luban}
\institute{
  \inst{1} Universit\"at Osnabr\"uck, Fachbereich Physik,
D-49069 Osnabr\"uck, Germany\\
  \inst{2} Ames Laboratory \& Department of Physics and Astronomy,
Iowa State University, Ames, Iowa 50011, USA
}
\pacs{75.10.Jm}{Quantized spin models}

\begin{document}

\maketitle

O. Waldmann \cite{Wal:EPL01} has shown that some spin systems,
which fulfill the condition of a weakly homogeneous coupling
matrix, have a spectrum whose minimal energies
$E_{\text{min}}(S)$, or maximal energies $E_{\text{max}}(S)$,
are rather poorly approximated by a quadratic dependence on the
total spin quantum number.  We comment on this observation and
provide the new argument that, under certain conditions, the
approximating parabolas appear as natural bounds of the spectrum
generated by spin coherent states.

In our article \cite{SSL:EPL01} we demonstrated that the spectrum
of interacting spin systems, with Hamiltonians of the following
form
$
\op{H}
\equiv
\sum_{\mu\nu}J_{\mu\nu} \op{\vec{s}}_\mu\cdot\op{\vec{s}}_\nu
,
$
is bounded by an upper and a lower parabola, i.e. all energy
eigenvalues lie between two curves which depend quadratically on
the total spin quantum number $S$. This proof is rigorous and
general and assumes only that the coupling constants
$J_{\mu\nu}$ satisfy
$
J_{\mu\nu}
=
J_{\nu\mu}
, 
J_{\mu\mu}=0
,
j\equiv \sum_\nu
J_{\mu\nu}
,
$
with $j$ being independent of $\mu$. The latter is a kind of
weak homogeneity assumption. In our notation $\op{\vec{s}}_\mu$
are single-spin operators of length $s$.

In the last section of our article \cite{SSL:EPL01} we also
reported that the bounding parabolas when shifted by an
appropriate amount can provide reasonable approximations
(``approximating parabolas") to the boundaries of the exact
energy spectrum for certain Heisenberg spin systems. Whereas the
bounding parabolas follow rigorously from the assumptions, we
made no such claim for the displaced ``approximating
parabolas". In fact, Waldmann \cite{Wal:EPL01} has provided
several interesting examples of Heisenberg spin systems where
the exact minimal energies $E_{\text{min}}(S)$, or maximal
energies $E_{\text{max}}(S)$, are not accurately approximated by
a quadratic dependence on the total spin quantum number $S$,
even though the bounding parabolas do apply for those systems
since the requirements are met. The observations of
\cite{Wal:EPL01} are valuable in the sense that they help to
clarify the conditions under which rotational bands, which are
rather frequently observed \cite{CCF:CEJ96,ACC:ICA00,ScL:PRB},
appear in magnetic systems.

In order to show that the approximating parabolas appear as
natural bounds of the spectrum generated by spin coherent
states, we first consider the convex set
\begin{equation}\label{4}
E_{\text{qm}}
\equiv
\left\{
\left(
\Tr{\op{H}\op{W}},
\Tr{\op{\vec{S}}^2\op{W}}
\right)
\Big|
\,
\op{W} \text{ statistical operator}
\right\}
\subset
\mathbb{R}^2
\ ,
\end{equation}
and its subset
\begin{equation}\label{5}
E_{\text{cl}}
=
\left\{
\left(
\bra{\Omega}\op{H}\ket{\Omega},
\bra{\Omega}\op{\vec{S}}^2\ket{\Omega},
\right)
\Big|
\,
\ket{\Omega} \text{ spin coherent product state}
\right\}
\subset
\mathbb{R}^2
\ .
\end{equation}
Here $\ket{\Omega}$ denotes any tensor product of $N$ spin
coherent states $\ket{\vec{\Omega}_i}, i=1, \dots, N$.  The
$\vec{\Omega}_i$ are vectors parameterizing each spin coherent
state by a unit vector pointing along the expected direction of
the spin, i.e.
$\bra{\vec{\Omega}_i}\op{\vec{s}}\ket{\vec{\Omega}_i}=s\,\vec{\Omega}_i$.
Note that for Heisenberg Hamiltonians
$
\bra{\Omega}\op{H}\ket{\Omega}
=
s^2 h(\Omega)
,
$
where $h(\Omega)$ denotes the classical Hamiltonian \cite{Simon}, whereas
$
\bra{\Omega}\op{\vec{S}}^2\ket{\Omega}
=
S_{\text{cl}}^2 + N\, s
\ ,\qquad
S_{\text{cl}}^2
=
s^2 \left(\sum_{i}\vec{\Omega}_i  \right)^2
.
$
Denote by
$\tilde{E}_{\text{min}}(S)$ the minimum of all $E$ for which
$(E,S(S+1))\in E_{\text{qm}}$. It is clear that
$\tilde{E}_{\text{min}}(S)\le E_{\text{min}}(S)$ , where
$E_{\text{min}}(S)$ denotes the minimal eigenvalue of $\op{H}$
within the subspace of total spin quantum number $S$.
Analogously,
$\tilde{E}_{\text{max}}(S)\ge E_{\text{max}}(S)$.

Using similar arguments as Berezin and Lieb
\cite{B,L} one can prove the following theorem \cite{Lieb:PC}.
\begin{theorem}\label{T1}
Let the Hamilton operator be of the given form given
and assume that
$
E_{\text{min}}^{\text{cl}}(S_{\text{cl}})
\ge
\frac{j-j_{\text{min}}}{N} S_{\text{cl}}^2 + j_{\text{min}} N
s^2
,
$
with $j_{\text{min}}$ being the minimal eigenvalue of the matrix
$(J_{\mu\nu})$.
Then for $S(S+1)\ge N\,s$ there holds
$
\tilde{E}_{\text{min}}(S)
\le
p_{\text{L}}(S)
\equiv
\frac{j-j_{\text{min}}}{N} S (S+1) + j_{\text{min}} (N\, s^2+s)-j\,s
,
$
where $p_{\text{L}}(S)$ coincides with the lower approximating parabola
of Ref.~\cite{SSL:EPL01}.
\end{theorem}

An analogous result can be proven which yields a lower bound for
$\tilde{E}_{\text{max}}(S)$.  The assumption made above is
fulfilled for a large class of spin systems, especially for
weakly homogeneous systems, including the examples given in
Refs.~\cite{SSL:EPL01,ScL:PRB}. This theorem, to some extent,
explains the good approximation of the true boundaries of the
spectrum by the shifted parabolas for those systems where
$\tilde{E}_{\text{min}}(S)$ and $E_{\text{min}}(S)$ are close
(analogously for the maxima). Theorem 1 would yield exact bounds
for the spectrum if the polygon $P$ with the vertices
$(E_{\text{min}}(S), S(S+1))$ and $(E_{\text{max}}(S), S(S+1))$
would circumscribe a convex figure, which then would be
identical with $E_{\text{qm}}$.  Actually the polygon $P$ seems
to be slightly concave for the examples we considered.

The interesting case (a) of \cite{Wal:EPL01} shows a lowest
level in the $(S=0)$-sector of the spectrum which is above the
approximating parabola and thus might appear to contradict our
findings. However, there is no contradiction since for such
small total spin quantum numbers the condition $S(S+1)\ge N\,s$
cannot be fulfilled and thus the true minimal energy may lie
above the approximating parabola.

Summarizing, our approximating parabolas appear as natural
bounds of the convex hull of the spectrum generated by spin
coherent states under the conditions specified by Theorem 1.
Therefore, these parabolas are of genuine classical origin, as
correctly pointed out by O.~Waldmann. But this does not mean
that they are necessarily poor approximations. It only shows
that in those cases where the approximations by parabolas are
reasonable the shape of the quantal spectrum is essentially
determined by classical physics.

\end{document}